\documentstyle[11pt,newpasp,twoside,epsf]{article}
\def\edcomment#1{\iffalse\marginpar{\raggedright\sl#1\/}\else\relax\fi}
\reversemarginpar

\begin{document}
\title{Yet MORE PM Survey Discoveries....?}

\author{A.J. Faulkner$^1$, M. Kramer$^1$, G. Hobbs$^1$, A.G. Lyne$^1$,
R.N. Manchester$^2$, F. Camilo$^3$, V.M. Kaspi$^5$, I.H. Stairs$^6$,
N. D'Amico$^4$, A. Possenti$^4$, D.R. Lorimer$^1$, M.A. McLaughlin$^1$}

\affil{$^1$University of Manchester, Jodrell Bank Observatory,  UK\\
$^2$ATNF, CSIRO, PO Box 76, Epping NSW 1710, Australia\\
$^3$Center for Space Research, MIT, Cambridge MA 02139, USA\\
$^4$Osservatorio Astronomico di Bologna, 40127 Bologna, Italy\\
$^5$Physics Dept., McGill University, Montreal, Quebec H3W 2C4, Canada\\
$^6$University of British Columbia, Vancouver, B.C. V6T 1Z1, Canada}

\begin{abstract}

The Parkes Multibeam (PM) survey has found nearly 50\% of the $>$1400
known pulsars and detected 20\% of the pulsars in binary systems (75\%
within the survey area), few of these are millisecond pulsars with
orbits of less than one day. It is known that the normal search
techniques used in the PM Survey have selected against finding binary
pulsars, due to the long integration times used. The relative
performance of the search techniques used and the improvements made
prior to the complete reprocessing of the data currently being
undertaken are discussed.  The processing is being done on COBRA - the
new supercomputer at Jodrell Bank. The results of the reprocessing, as
well as adding more solitary pulsars, will give more confidence on the
population and types of binaries in the Galactic plane. However,
it can be shown that for millisecond pulsars in binary orbits of
between 20min and a few hours, the search techniques are still
relatively insensitive.

\end{abstract}

\section{Introduction}

The Parkes Multibeam Survey (PM Survey) has discovered nearly half the
known pulsars. Observations started in 1997 and completed in 2002. The
PM survey used the 13-beam receiver at the 64-metre Parkes Radio
Telescope and covered a strip along the Galactic plane with $|b|<5^o$
and $l=260^o$ to $l=50^o$. The observations were carried out with a
0.25ms sampling time, collecting $2^{23}$ samples, in an approximately
35-min observation. The nominal limiting flux density is about 0.2mJy,
for dispersion measures (DM) of less than 300 cm$^{-3}$pc. (Further
details of the PM system can be found in Manchester et
al. (2001)). There are approximately 2670 pointings, each of 13 beams,
making a total of 34,710 integrations. The data are stored on some 154
DLT tapes for off-line processing.

Since the start of the survey, improved search techniques have been
implemented and there has been some changes to the interference
filters used. Consequently, there are more pulsars expected to be
discovered using the existing data.  With the increased computing
power now available, particularly in the form of beowulf clusters such
as Cobra at Jodrell Bank, it has been decided to reprocess the entire
data set.

\section{Binary Pulsar Discoveries}

\begin{figure}[!t]
\begin{center}
\plotfiddle{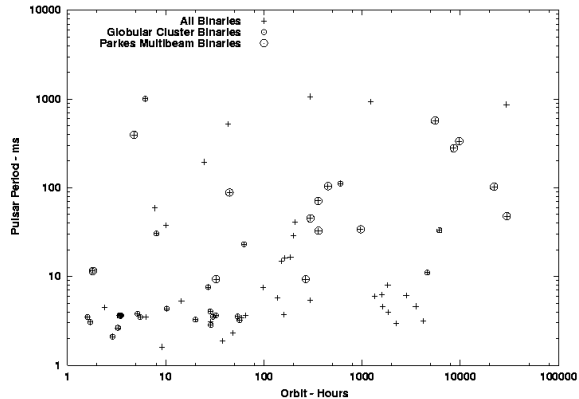}{1.7in}{-90.}{30.}{30.}{-130}{155}
\end{center}

\caption{\itshape All binary pulsars, showing the relationship betweem
pulsar period and binary orbit.}  \normalfont
\end{figure}

There have been 73 binary pulsars discovered, of which 15 have been
discovered or re-detected in the PM survey (there are 19 known in the
survey area). There is a large number, 25, which have been discovered
in searches of Globular Clusters (GC). The reasons for being so many
GC binary pulsars discovered is that as well as having  a lot of stars
in a restricted space, so stars often ``capture'' each other, the
whole of the GC is in a narrow range of dispersion measures. It is,
therefore, feasible to use more computationally expensive acceleration
searches. This illustrates the benefit of using improved acceleration
search algorithms with long integrations as discussed in Friere (2000)
and Camilo et al. (2000). 

The plot, in figure 1, of pulsar periods and orbits for these systems
highlights that PM binary pulsars may have some selection
effects. There are four binary pulsars which are within the PM survey
area and not reported as detected. Two of the undetected pulsars
B1855+09 and B1913+16 have flux levels of 4mJy (DM 13.3) and 0.7mJy
(DM 169) respectively at 1400MHz, well above the minimum detectability
levels. The other two do not have flux levels recorded for 1400MHz,
but have DM's and periods where the PM survey is reasonably sensitive.

Considering B1913+16, for example, it is a  pulsar with a
period of 59ms, but with a fast orbital period of 7.7hours. The full
FFT would suffer significant smearing of upto 10 Fourier bins, thus
limiting sensitivity.

Also, it can be seen that, apart from B1744-24A (a pulsar in GC
Terzian 5) and J1141-6545 which are both very bright, there are no
pulsars detected with orbits of less than 1 day, which could be
partially due to selection effects. There are no binary pulsars with
periods of less than 9-ms, however, there is a group with medium -
long periods (50 - 500ms) and long orbits ($>$230 days), these can be
detected using non-accelerated searches.

There is a clear possibility of finding further binary pulsars in the
PM data.

\section{PM Reprocessing Search algorithms}

The PM survey has used a long integration of 35 mins for high
sensitivity. However, the integration time makes it insensitive to
fast binary pulsars due to smearing of received power across a number
of Fourier bins. The processing to date has almost all only used a
frequency domain search of the full data set for each
observation. Consequently, the search has had the predicted
sensitivities only for solitary pulsars up to about 2-sec period. The
reprocessing of the data includes searches for accelerated (binary)
pulsars, long period pulsars and single pulses - looking for ``Giant
Pulses'' at a particular DM (McLaughlin 2001).

\begin{itemize}
\item \bf Frequency Domain Search  \normalfont -
Most pulsars found have periods of one second or less and are
effectively searched for using the frequency domain. This is due to
the great efficiency of the FFT algorithm at separating many different
frequencies concurrently.

This is the same search as performed previously, using the full length
FFT. However, improvements in the ``birdie filters'' means that less
of the spectrum is removed.

\item \bf Time Domain Search \normalfont -
Long period pulsars ( $>$ 2 sec period) are most sensitively searched for
using a time domain search, efficiently done by using a Fast
Folding Algorithm (FFA) (Staelin 1969). This is because at low
frequencies the separation of periods becomes relatively coarse using
an FFT. By only searching a limited period range the FFA is reasonably
efficient computationally.

\item \bf Linear Acceleration Search \normalfont - An efficient
technique to search for linearly accelerating pulsars has been written
and tested (e.g Anderson 1992), but has not been used on large
quantities of data. Known as a ``stack search'', this operates by the
relative shifting and summing of independent partial FFT's of the data
(in this case 16 segments). It is very efficient in that many
accelerations can be tested with only one set of transformed
segments. Its limitations are that there is a loss of approximately
20\% in sensitivity (see tests later and Anderson 1992) over a fully
coherent search and only pulsars which are in orbits long enough to be
accelerating linearly over the time of the observation, will have the
maximum sensitivity.

\item \bf Phase Acceleration Search \normalfont -
This approach searches for periodicities in the variations of the
observed pulsar period, using the frequency domain - an FFT of an FFT
(Ransom 2001). This is capable of identifying pulsars which have
completed more than 1.5 orbits over the period of the observation. In
the PM survey these would be very fast binaries, having orbits less
than 20-mins. Such binaries do have been detected in X-rays,
for example X1820-303 (Stella, Priedorsky \& White 1987, which has a
685-sec orbit.

\end{itemize}

\section{Search algorithm testing}

\begin{figure}[!t]
\begin{center}
\plottwo {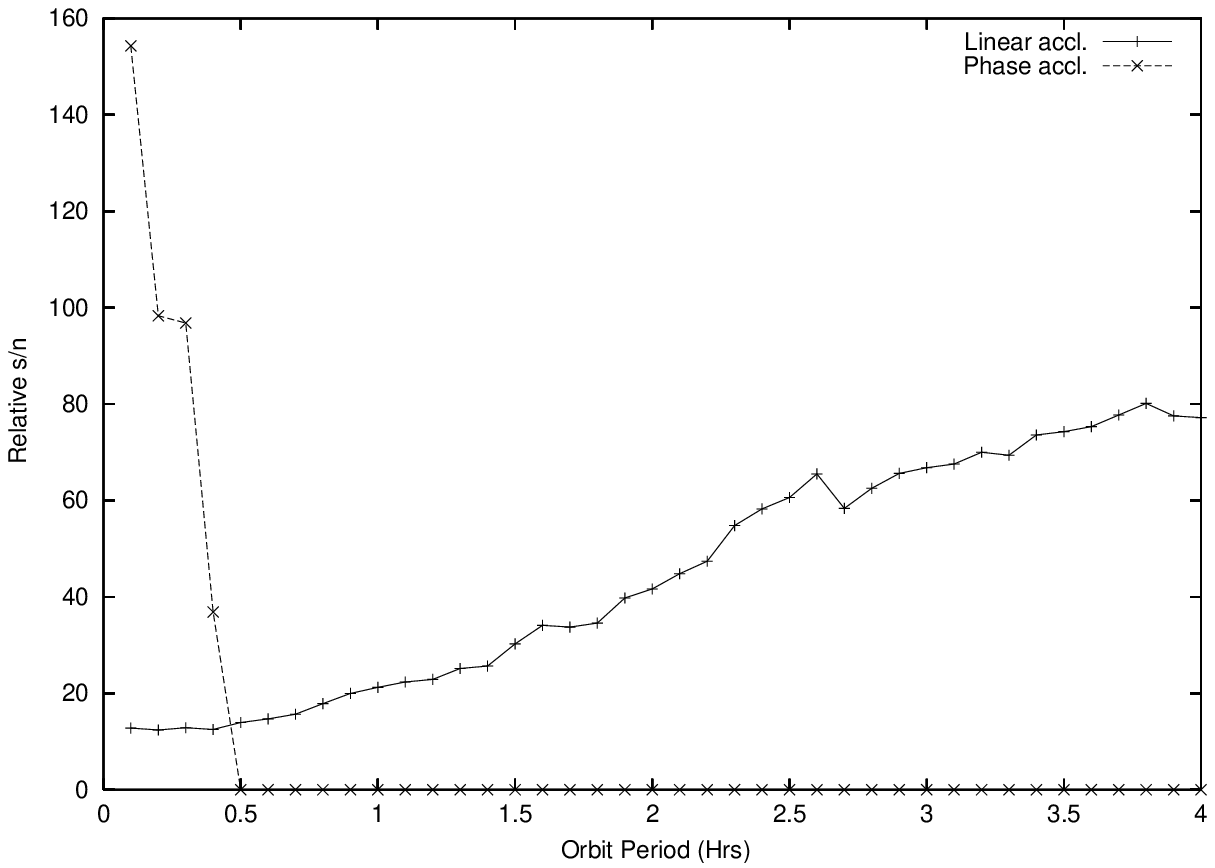}{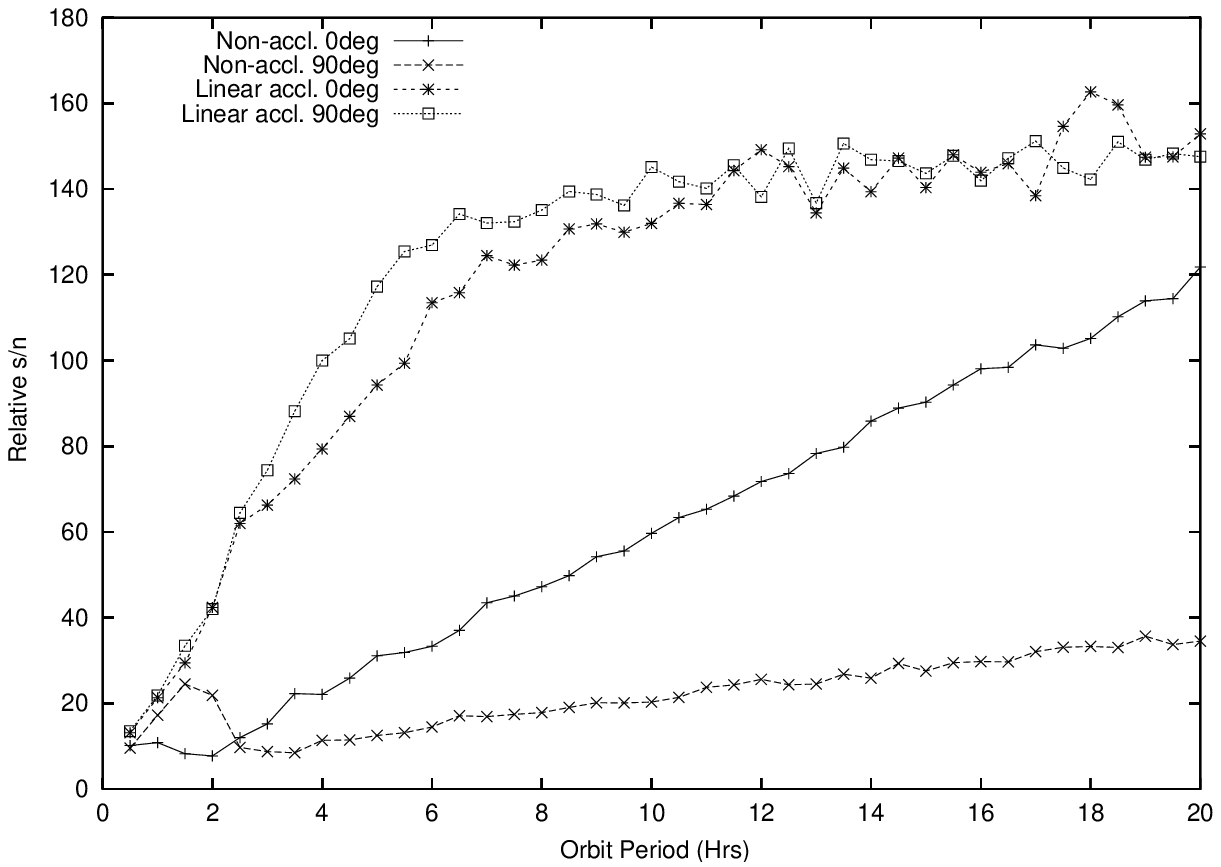}
\end{center}
\caption{\itshape Sensitivity to pulsars in binary orbits for the
phase acceleration, ``stack'' linear accl. searches and unaccelerated
searches. In each case the pulsar is in a circular orbit with a period
of 10ms. The sampling time is .25ms and has a 35min integration. Note
the difference in sensitivity for starting at different phase angles
of the orbit. For reference, the sensitivity to an isolated pulsar is
215 for a non-accel. search and 168 for a linear accl. search.}
\normalfont
\end{figure}

Figure 2 shows the sensitivity of the search algorithms to simulated
pulsars with various parameters. Although the segmented search
significantly increases sensitivity to relatively fast binaries,
it can clearly be seen that there is still a region of low sensitivity
for binary pulsars between 20min and 6 hours.

\section{Cobra Reprocessing}

Reprocessing of data at Jodrell Bank will be on Cobra - a cluster of
91 processor boards incorporating 182 processors. A full tape takes
approximately 1 day to process, while using 100 processors. It is
anticipated that a significant number of new pulsars will be
discovered in the PM Survey data; most will be normal solitary
pulsars, however, there \itshape may \normalfont be some really
exciting objects.

\end{document}